\documentclass[aps,twocolumn,prb,eqsecnum,showpacs]{revtex4}
\usepackage{graphicx}
\begin{document}
\draft
\preprint{\today}
\title{Modified Spin-Wave Theory of Nuclear Magnetic Relaxation in
       One-Dimensional Quantum Ferrimagnets:
       Three-Magnon versus Raman Processes}
\author{Shoji Yamamoto and Hiromitsu Hori}
\address{Division of Physics, Hokkaido University,
         Sapporo 060-0810, Japan}
\date{\today}
\begin{abstract}
Nuclear spin-lattice relaxation in one-dimensional Heisenberg ferrimagnets
is studied by means of a modified spin-wave theory.
Calculating beyond the first-order mechanism, where a nuclear spin
directly interacts with spin waves through the hyperfine coupling, we
demonstrate that the exchange-scattering-enhanced three-magnon nuclear
relaxation may generally predominate over the Raman one with increasing
temperature and decreasing field.
Recent proton spin-lattice relaxation-time ($T_1$) measurements on
the ferrimagnetic chain compound
NiCu(C$_7$H$_6$N$_2$O$_6$)(H$_2$O)$_3$$\cdot$2H$_2$O
suggest that the major contribution to $1/T_1$ be made by the three-magnon
scattering.
\end{abstract}
\maketitle

\section{Introduction}

   Nuclear magnetic resonance (NMR) is an effective probe to the
collective motions of electronic spins and therefore we take a great
interest in microscopically interpreting it.
The spin-wave formalism has played a crucial role in this context.
Van Kranendonk, Bloom \cite{K545} and Moriya \cite{M23,M641} made their
pioneering attempts to describe the nuclear spin-lattice relaxation time
$T_1$ in terms of spin waves.
Oguchi and Keffer \cite{O405} further developed the spin-wave analysis
considering the three-magnon nuclear relaxation mechanism as well as the
Raman one.
Pincus and Beeman \cite{P398,B359} claimed that the exchange correlation
between spin waves should significantly accelerate the nuclear spin
relaxation.
The spin-wave excitation energy is usually much larger than the nuclear
resonance frequency and thus the single-magnon relaxation process is
rarely of significance.
The Raman process consequently plays a leading role in the nuclear
spin-lattice relaxation.
Because of the $(4S)^{-1}$-damping factor to the Holstein-Primakoff magnon
series expansion, the multi-magnon scattering is much less contributive
within the first-order process, where a nuclear spin directly interacts
with spin waves through the hyperfine coupling.
However, the second-order process, where a nuclear spin flip induces
virtual spin waves which are then scattered thermally via the four-magnon
exchange interaction, may generally enhance the relaxation rate.
This is the fascinating scenario written by Pincus and Beeman.

   It is unfortunate that their spin-wave nuclear relaxation theory is not
effective in low dimensions but is valid far below the transition
temperature.
The conventional spin-wave theory applied to low-dimensional magnets ends
in failure with diverging magnetizations.
In such circumstances, Takahashi \cite{T168} gave a fine description of
the low-dimensional ferromagnetic thermodynamics in terms of modified spin
waves.
His idea of introducing a constraint on the magnetization so as to control
the number of spin waves was further applied to antiferromagnets
\cite{T2494,H4769,T5000} and ferrimagnets.\cite{Y14008,Y11033,O8067}
Even frustrated antiferromagnets \cite{H2887,C7832,D13821} and random-bond
ferromagnets \cite{W014429} were discussed within this renewed spin-wave
scheme.
\begin{figure}[b]
\centering
\includegraphics[width=72mm]{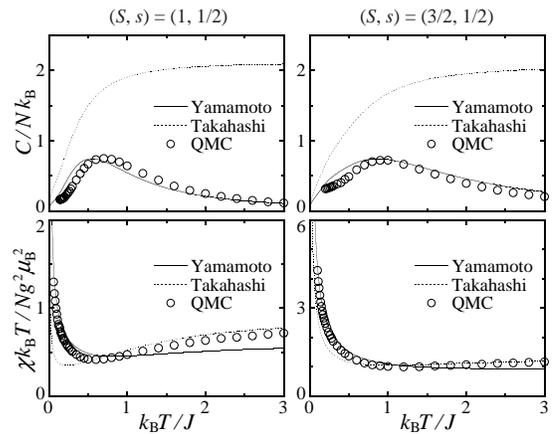}
\caption{Modified spin-wave calculations of the specific heat $C$ and
         the magnetic susceptibility $\chi$ as functions of temperature
         for the ferrimagnetic Heisenberg chains.
         The original (Takahashi) and our (Yamamoto) schemes are
         compared with quantum Monte Carlo (QMC) calculations.}
\label{F:MSWdemo}
\end{figure}
\begin{figure*}
\centering
\includegraphics[width=144mm]{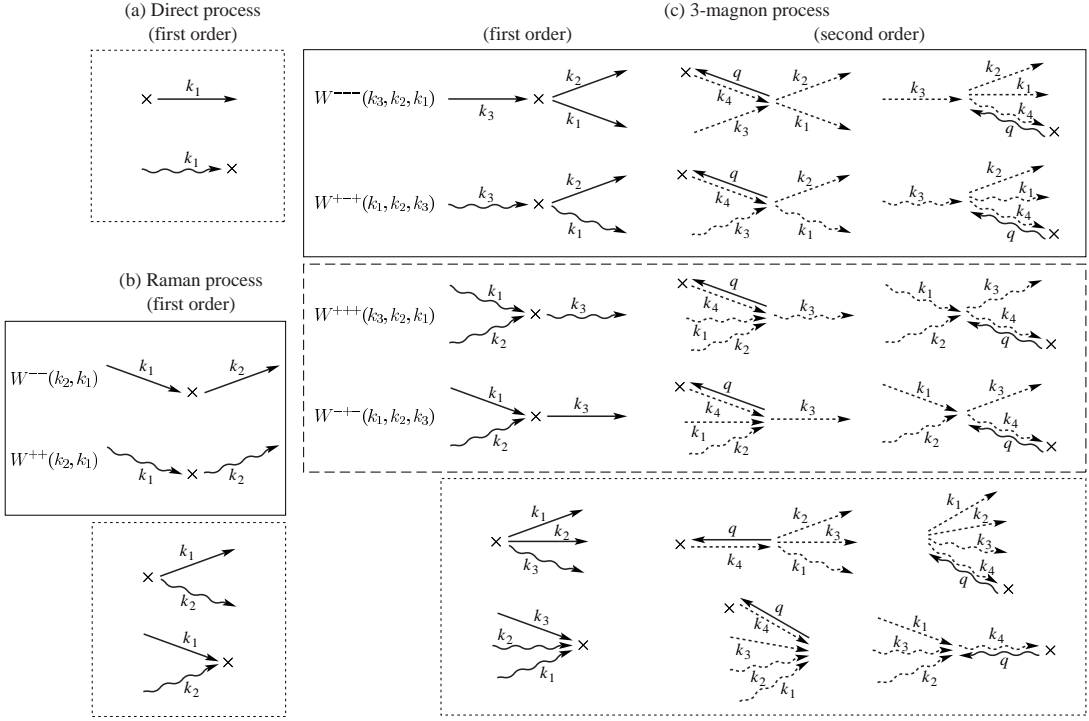}
\caption{Diagrammatic representation of various nuclear spin-lattice
         relaxation processes.
         Solid arrows, designating spin waves which are emitted in the
         first-order processes, induce a nuclear spin flip ($\times$) via
         the hyperfine interaction, while broken arrows, depicting the
         four-magnon exchange correlations, thermally scatter the
         first-order spin waves as {\it virtual excitations}, where spin
         waves of ferromagnetic and antiferromagnetic aspect are
         distinguishably drawn by straight and wavy arrows, respectively.
         (a) The first-order direct (single-magnon) relaxation processes;
         (b) The first-order Raman (two-magnon) relaxation processes;
         (c) The first-order and second-order three-magnon relaxation
             processes, where $q=-k_4$, are related to
             each other through nonlinear equations and are therefore
             inseparable.
         Considering the nuclear-electronic energy conservation, processes
         in solid and dotted frames are of great and little significance,
         respectively, whereas those in broken frames are relevant
         according to the constituent spins $S$ and $s$.
         Labels $W^{\sigma\sigma}$ and $W^{\sigma\sigma'\sigma}$ on
         feasible processes are explained in Appendix \ref{A:T1}.}
\label{F:diagram}
\end{figure*}

   The ferrimagnetic modified spin-wave theory is particularly useful in
illuminating both static \cite{O8067,Y1024,N214418} and dynamic
\cite{Y157603,H054409,H1453} properties.
One-dimensional ferrimagnets have lately attracted much attention
especially in the context of designing molecule-based
ferromagnets.
Assembling molecular bricks in such a way as to obtain a low-dimensional
system with a nonzero resultant spin in the ground state and then coupling
the chains or the layers again in a ferromagnetic fashion, one can in
principle obtain a molecular magnet.
A series of bimetallic chain compounds \cite{K782,K3325} were synthesized
in such a strategy.
Another approach \cite{C1976,C2940} to molecular magnets consists of
bringing into interaction metal ions and stable organic radicals.
Homometallic materials such as tetrameric bond-alternating chain compounds
\cite{E4466} and trimeric intertwining double-chain compounds \cite{D83}
are distinct ferrimagnets of topological origin.
Such synthetic endeavors have stimulated several experimentalists
\cite{F8410,F1073,F433} to measure $T_1$ on ferrimagnetic chain compounds.
Thus motivated, we have started a modified spin-wave exploration
\cite{H1453} of the nuclear spin dynamics in one-dimensional Heisenberg
ferrimagnets beyond the conventional Raman mechanism.
Here we give a full description of the theory and finally show a strong
evidence of {\it the proton spin relaxation in the ferrimagnetic chain
compound $NiCu(C_7H_6N_2O_6)(H_2O)_3\cdot 2H_2O$
being mediated by the three-magnon scattering rather than the Raman one}.

\section{Modified Spin-Wave Scheme}

   We consider ferrimagnetic Heisenberg chains of alternating spins $S$
and $s$, as described by the Hamiltonian
\begin{equation}
   {\cal H}
      =\sum_{n=1}^N
       \big[
        J\mbox{\boldmath$S$}_{n}\cdot
         (\mbox{\boldmath$s$}_{n-1}+\mbox{\boldmath$s$}_{n})
       -(g_SS_n^z+g_ss_n^z)\mu_{\rm B}H
       \big].
   \label{E:H}
\end{equation}
Introducing bosonic operators for the spin deviation in each sublattice
via
$S_i^+=(2S-a_{i}^\dagger a_{i})^{1/2}a_{i}$,
$S_i^z=S-a_{i}^\dagger a_{i}$,
$s_i^+=b_{i}^\dagger(2s-b_{i}^\dagger b_{i})^{1/2}$,
$s_i^z=-s+b_{i}^\dagger b_{i}$,
and assuming that $O(S)=O(s)$, we expand the Hamiltonian with respect to
$1/S$ as
${\cal H}=-2SsJN+{\cal H}_1+{\cal H}_0+O(S^{-1})$, where ${\cal H}_i$
contains the $O(S^i)$ terms which are explicitly given in
Appendix \ref{A:H}.

   Our scheme \cite{Y064426} of modifying the spin-wave theory is distinct
from the original idea proposed by Takahashi \cite{T2494} and
Hirsch {\it et al.}.\cite{H4769}
In their way of suppressing the divergence of the sublattice
magnetizations, an effective Hamiltonian with a Lagrange multiplier
included, instead of the original Hamiltonian, is diagonalized subject to
zero staggered magnetization.
On the other hand, we first diagonalize the Hamiltonian keeping the
dispersion relations free from temperature and then introduce a Lagrange
multiplier in order to minimize the free energy subject to zero staggered
magnetization.
The two approaches are compared in Fig. \ref{F:MSWdemo}.
Our new scheme is much better at describing the antiferromagnetically
peaked specific heat and the low-temperature diverging susceptibility.
Our approach converges into the paramagnetic behavior at high temperatures
for both the specific heat and susceptibility.

\section{Nuclear Spin-Lattice Relaxation}
\begin{figure*}
\centering
\includegraphics[width=144mm]{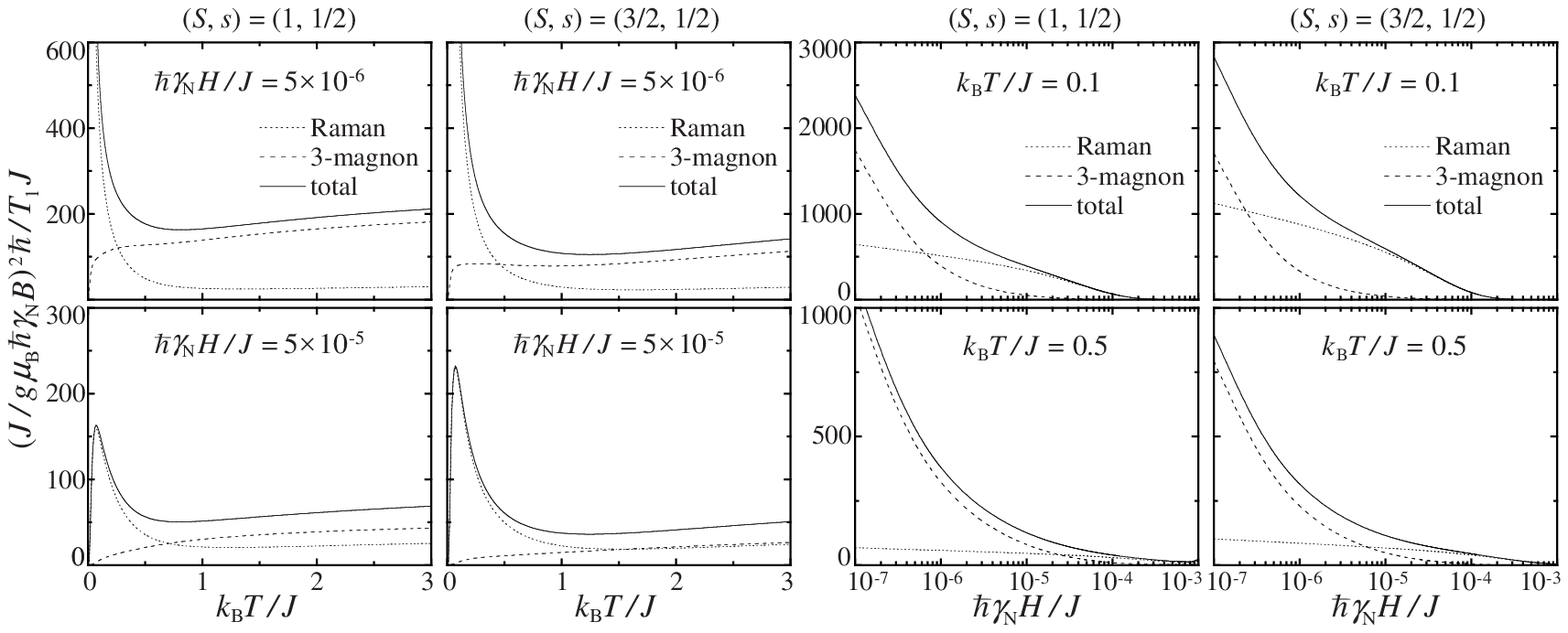}
\caption{Modified spin-wave calculations of typical temperature (the left
         four) and field (the right four) dependences of the nuclear
         spin-lattice relaxation rate, where $g_S=g_s\equiv g$,
         $A^\tau/B^\tau=0$ and $(B^-/B^z)^2=4$.
         $1/T_1^{(2)}$ and $1/T_1^{(3)}$ are plotted by dotted and broken
         lines, respectively, while $1/T_1^{(2)}+1/T_1^{(3)}\equiv 1/T_1$,
         which is observable, by solid lines.}
\label{F:T1}
\end{figure*}

   The hyperfine interaction is generally expressed as
\begin{eqnarray}
   &&
   {\cal H}_{\rm hf}
   =g_S\mu_{\rm B}\hbar\gamma_{\rm N}I^+
    {\textstyle \sum_n}
     \big(
      {\textstyle\frac{1}{2}}A_n^-S_n^-+A_n^zS_n^z
     \big)
   \nonumber\\
   &&\qquad\ 
   +g_s\mu_{\rm B}\hbar\gamma_{\rm N}I^+
    {\textstyle \sum_n}
     \big(
      {\textstyle\frac{1}{2}}B_n^-s_n^-+B_n^zs_n^z
     \big),
\end{eqnarray}
where $A_n^\sigma$ ($B_n^\sigma$) is the dipolar coupling tensor
between the nuclear and $n$th larger (smaller) electronic spins.
Since ${\cal H}_0$ and ${\cal H}_{\rm hf}$ are both much smaller than
${\cal H}_1$, they act as perturbative interactions to the linear
spin-wave system.
If we consider up to the second-order perturbation with respect to
${\cal V}\equiv{\cal H}_0+{\cal H}_{\rm hf}$, the probability of a nuclear
spin being scattered from the state of $I^z=m$ to that of $I^z=m+1$ is
given by
\begin{equation}
   W=\frac{2\pi}{\hbar}\sum_f
     \Biggl|\Bigl\langle f\Bigl|
      {\cal V}+\sum_{m(\neq i)}
      \frac{{\cal V}|m\rangle\langle m|{\cal V}}{E_i-E_m}
     \Bigr|i\Bigr\rangle\Biggr|^2
     \delta(E_i-E_f),
   \label{E:W}
\end{equation}
where $i$ and $f$ designate the initial and final states of the
unperturbed electronic-nuclear spin system, whose energies are
$E_i$ and $E_f$, respectively.
The nuclear spin-lattice relaxation time is then given by
$T_1=(I-m)(I+m+1)/2W$.

   Equation (\ref{E:W}) contains various scattering processes, which are
diagrammatically shown in Fig. \ref{F:diagram}.
Due to the considerable difference between the nuclear and electronic
energy scales, $\hbar\omega_{\rm N}\ll J$, the direct process, involving a
single spin wave, is rarely of significance.
Considering further that the antiferromagnetic spin waves are higher in
energy than the ferromagnetic ones, $\omega_k^-<\omega_k^+$, at moderate
fields, the intraband spin-wave scattering dominates the Raman relaxation
rate $1/T_1^{(2)}$, whereas both the intraband and interband spin-wave
scatterings contribute to the three-magnon relaxation rate $1/T_1^{(3)}$.
Within the first-order mechanism, $1/T_1^{(3)}$ is much
smaller than $1/T_1^{(2)}$.\cite{O405}
However, the first-order relaxation rate may be enhanced through
the second-order mechanism.
We consider the leading second-order relaxation, that is, the
exchange-scattering-induced three-magnon processes, as well as the
first-order ones.
The second-order Raman processes, containing two virtual magnons,
are much more accidental due to the momentum conservation and much less
contributive due to the $(4S)^{-1}$-damping factor in the
Holstein-Primakoff magnon series expansion.
As for the four-magnon scattering, the first-order processes are
nonexistent, whereas the second-order ones originate in the six-magnon
exchange interaction and therefore contain two virtual magnons.
Thus and thus, all other higher-order processes do not significantly
change the relaxation scenario.
We explicitly formulate $1/T_1^{(2)}$ and $1/T_1^{(3)}$ in Appendix
\ref{A:T1}.

\section{Three-Magnon versus Raman Processes}
\begin{figure}[b]
\centering
\includegraphics[width=72mm]{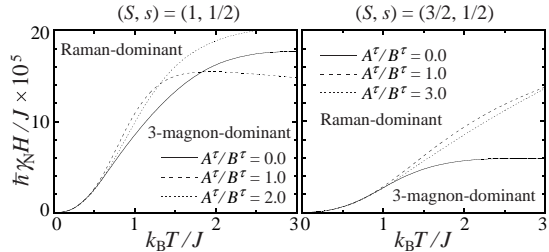}
\caption{The crossover point as a function of temperature and an applied
         field, where $g_S=g_s$ and $(B^-/B^z)^2=4$.}
\label{F:PhD}
\end{figure}
\begin{figure*}
\centering
\includegraphics[width=144mm]{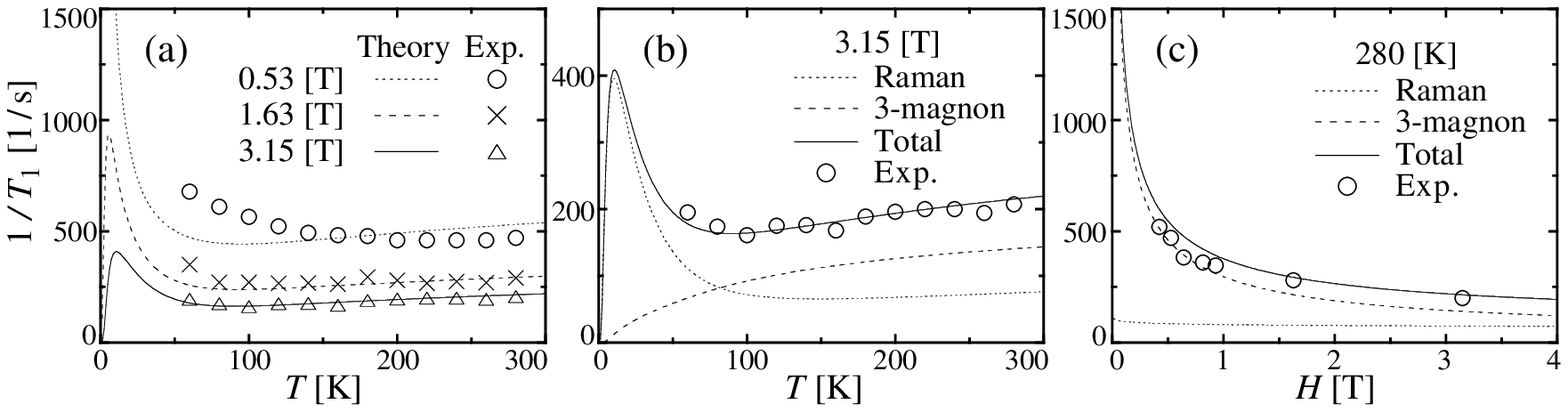}
\caption{Proton spin relaxation-time measurements on
         NiCu(pba)(H$_2$O)$_3$$\cdot$2H$_2$O (Ref. \onlinecite{F433}) compared
         with our theory.
         (a) $1/T_1$ as a function of temperature at various values of
             an applied field;
         (b) $1/T_1$ as a function of temperature at $3.15\,\mbox{T}$;
         (c) $1/T_1$ as a function of an applied field at $280\,\mbox{K}$.
             In (b) and (c), the Raman ($1/T_1^{(2)}$) and three-magnon
             ($1/T_1^{(3)}$) contributions are also plotted by dotted and
             broken lines, respectively.}
\label{F:experiment}
\end{figure*}

   We calculate the cases of $(S,s)=(1,\frac{1}{2})$ and
$(S,s)=(\frac{3}{2},\frac{1}{2})$, which are relevant to several major
materials.\cite{C1976,D83,K3325}
Figure \ref{F:T1} shows $1/T_1$ as a function of temperature and an
applied field.
The exchange-scattering-enhanced three-magnon relaxation rate generally
grows into a major contribution to $1/T_1$ with increasing temperature and
decreasing field.
As temperature increases, $\bar{n}_k^-$ decreases at $k\simeq 0$ but
otherwise increases.\cite{Y2324}
In one dimension, excitations at $k\simeq 0$ predominate in the Raman
processes, while all the excitations are effective in the three-magnon
processes.
$1/T_1^{(2)}$ and $1/T_1^{(3)}$ are hence decreasing and increasing
functions of temperature, respectively, unless temperature is so high as
to activate the antiferromagnetic spin waves.
The field dependences of $1/T_1^{(2)}$ and $1/T_1^{(3)}$ are also in
striking contrast.
At moderately low temperatures and weak fields,
$\hbar\omega_{\rm N}\ll k_{\rm B}T\ll J$, $T_1^{(2)}$ is approximately
evaluated as
\begin{eqnarray}
   &&
   \frac{1}{T_1^{(2)}}
   \simeq
   \frac{2[(g_SA^zS-g_sB^zs)\mu_{\rm B}\hbar\gamma_{\rm N}]^2}
        {\pi\hbar Ss(S-s)J}
   \nonumber \\
   &&\qquad\times
   {\rm exp}
   \Bigl[
    -\frac{(g_S+g_s)\mu_{\rm B}H}{2k_{\rm B}T}
   \Bigr]
   K_0\Bigl(\frac{\hbar\omega_{\rm N}}{2k_{\rm B}T}\Bigr),
\end{eqnarray}
where $K_0$ is the modified Bessel function of the second kind and
behaves as
$K_0(\hbar\omega_{\rm N}/2k_{\rm B}T)
 \simeq 0.80908-{\rm ln}(\hbar\omega_{\rm N}/k_{\rm B}T)$.
Thus the field dependence of $1/T_1^{(2)}$ is initially logarithmic
and then turns exponential with increasing field.
Equation (\ref{E:T1(3)}) is much less analytical but suggests much
stronger power-law diverging behavior with decreasing field.
Therefore, the three-magnon relaxation predominates over the Raman
one at weak fields.

   In Fig. \ref{F:PhD} we plot {\it the crossover points} on which
$1/T_1^{(2)}=1/T_1^{(3)}$.
A Raman-to-three-magnon crossover may generally be detected with
increasing temperature and decreasing field.
The ferrimagnetic nuclear spin-lattice relaxation is sensitive to another
adjustable parameter $A^\tau/B^\tau$, that is, the location of the probe
nuclei.\cite{Y2324}
At the special location of
$A^\tau/B^\tau\sim(d_s/d_S)^3\simeq(S/s)^\sigma$, where $d_S$ ($d_s$) is
the distance between the nuclear and larger (smaller) electronic spins,
the $\sigma$ excitation mode is almost invisible to the nuclear
spin.\cite{Y2324}
In the case of $(S,s)=(1,\frac{1}{2})$, for exmaple, the nuclear spin
located as $A^\tau/B^\tau\simeq 1/2$ hardly relaxes.
Any $T_1$ measurements should be performed away from such magic points.

\section{Interpretation of Experiments}

   We are further excited to compare our theory with recent experimental
findings.
Fujiwara and Hagiwara \cite{F433} measured $T_1$ for proton nuclei in the
bimetallic chain compound
NiCu(pba)(H$_2$O)$_3$$\cdot$$2$H$_2$O
($\mbox{pba}
 =1,3\mbox{-propylenebis(oxamato)}
 =\mbox{C}_7\mbox{H}_6\mbox{N}_2\mbox{O}_6$) \cite{P138}
comprising ferrimagnetic chains with alternating octahedral Ni$^{2+}$ and
square-pyramidal Cu$^{2+}$ ions bridged by oxamato groups.
The measured susceptibility \cite{H2209} is well reproduced with
$g_S=2.22$, $g_s=2.09$ and $J/k_{\rm B}\simeq 121\,\mbox{K}$.
Since the protons relevant to the $T_1$ findings are located in close
vicinity to Cu ions, we may set the coupling constants for
$A^\tau/B^\tau=0$.
We further assume that $B^z=1.37\times 10^{20}\,\mbox{T}^2/\mbox{J}$
and $(B^-/B^z)^2=5$, which can be consistent with the crystalline
structure.\cite{P138}

   The thus-calculated $1/T_1$ is compared with the observations in
Fig. \ref{F:experiment}.
Considering that there may be larger uncertainty in the experimental
analysis at lower temperatures,\cite{F433}
the theoretical and experimental findings are in good agreement and the
slight discrepancy between them may be attributable, for instance, to weak
momentum dependence of $B_k^\tau$ and the protons of wide distribution.
Figure \ref{F:experiment} (b) shows that {\it the increasing behavior of
$1/T_1$ at high temperatures originates from the three-magnon
contribution}.
Figure \ref{F:experiment}(c) more impressively demonstrates the
relevance of the three-magnon scattering to the proton spin relaxation.
{\it The strong field dependence can never be explained by the Raman
process}.
Since $1/T_1^{(3)}$ within the first-order mechanism stays much smaller
than the observations, {\it the exchange-scattering-induced three-magnon
process is essential in interpreting such accelerated relaxation}.
We are eager to have reliable observations at lower temperatures and
weaker fields.
More extensive NMR measurements on the related compounds are encouraged.

\section{Summary}

   There exist pioneering $T_1$ measurements on the layered ferromagnet
CrCl$_3$ \cite{N354} and the coupled-chain antiferromagnet
CsMnCl$_3\cdot$2H$_2$O,\cite{N5325} which give evidence of the relevant
three-magnon scattering.
However, they are both, in some sense, {\it classical} findings under the
existing three-dimensional long-range order.
No author has explored one-dimensional {\it quantum} ferrimagnetic
dynamics with particular interest in multi-magnon scattering beyond the
Raman mechanism.
We have reported {\it the first evidence of the three-magnon
scattering dominating one-dimensional nuclear spin relaxation}.
We hope the present research will stimulate further measurements and lead
to close collaboration between theoretical and experimental
investigations.

\acknowledgments
   The authors are grateful to T. Goto, N. Fujiwara and Y. Furukawa for
valuable comments.
This work was supported by the Ministry of Education, Culture, Sports,
Science and Technology of Japan, and the Iketani Science and Technology
Foundation.

\begin{widetext}
\begin{appendix}
\section{Spin-Wave Hamiltonian}\label{A:H}

   Performing the Fourier and then Bogoliubov transformations as
$N^{-1/2}\sum_n{\rm e}^{ {\rm i}k(n-1/4)}a_n^\dagger
=\alpha_k^\dagger{\rm ch}\theta_k-\beta_k{\rm sh}\theta_k$ and
$N^{-1/2}\sum_n{\rm e}^{-{\rm i}k(n+1/4)}b_n^\dagger
=\beta_k^\dagger{\rm ch}\theta_k-\alpha_k{\rm sh}\theta_k$,
where we abbreviate ${\rm cosh}\theta_k$ and ${\rm sinh}\theta_k$ as
${\rm ch}\theta_k$ and ${\rm sh}\theta_k$, respectively,
the Hamiltonian is represented as
\begin{eqnarray}
   &&
   {\cal H}_1
    =-2JN\bigl[2\sqrt{Ss}{\mit\Gamma}-(S+s){\mit\Lambda}\bigr]
     -\bigl[
       g_SS-g_ss-(g_S-g_s){\mit\Lambda}
      \bigr]\mu_{\rm B}HN
   \nonumber\\
   &&\qquad
     +J\sum_k
       \bigl[\omega_1^-(k)\alpha_k^\dagger\alpha_k
            +\omega_1^+(k)\beta_k^\dagger\beta_k
            +\gamma_1(k)(\alpha_k^\dagger\beta_k^\dagger+\alpha_k\beta_k)
       \bigl],
   \\
   &&
   {\cal H}_0
    =-2JN
      \Bigr[
       {\mit\Gamma}^2+{\mit\Lambda}^2
      -\Bigl(\sqrt{\frac{S}{s}}+\sqrt{\frac{s}{S}}\Bigr)
       {\mit\Gamma\Lambda}
      \Bigr]
     +J\sum_k
       \bigl[
        \omega_0^-(k)\alpha_k^\dagger\alpha_k
       +\omega_0^+(k)\beta_k^\dagger\beta_k
       +\gamma_1(k)(\alpha_k^\dagger\beta_k^\dagger+\alpha_k\beta_k)
       \bigr]
   \nonumber\\
   &&\qquad
     -\frac{J}{4N}\sum_{k_1,\cdots,k_4}\delta(k_1-k_2-k_3+k_4)
      \Bigl[
       V_0^{(1)}(k_1,k_2,k_3,k_4)
       \alpha_{k_1}^\dagger\alpha_{k_4}^\dagger
       \alpha_{k_2}\alpha_{k_3}
      +V_0^{(2)}(k_1,k_2,k_3,k_4)
       \beta_{k_1}^\dagger\beta_{k_4}^\dagger
       \beta_{k_2}\beta_{k_3}
   \nonumber\\
   &&\qquad
      +V_0^{(3)}(k_1,k_2,k_3,k_4)
       \alpha_{k_1}^\dagger\beta_{k_2}^\dagger
       \alpha_{k_3}\beta_{k_4}
      -V_0^{(4)}(k_1,k_2,k_3,k_4)
       (\alpha_{k_1}^\dagger\alpha_{k_2}\alpha_{k_3}\beta_{k_4}
       +{\rm H.c.})
   \nonumber\\
   &&\qquad
      -V_0^{(5)}(k_1,k_2,k_3,k_4)
       (\beta_{k_1}^\dagger\beta_{k_2}\beta_{k_3}\alpha_{k_4}
       +{\rm H.c.})
      +V_0^{(6)}(k_1,k_2,k_3,k_4)
       (\beta_{k_1}\alpha_{k_2}\beta_{k_3}\alpha_{k_4}
       +{\rm H.c.})
      \Bigr],
\end{eqnarray}
where
\begin{eqnarray}
   &&
   {\mit\Gamma}
    =\frac{1}{2N}\sum_k
     {\rm cos}\frac{k}{2}{\rm sh}2\theta_k,\ \ 
   {\mit\Lambda}
    =\frac{1}{2N}\sum_k({\rm ch}2\theta_k-1),
   \\
   &&
   \omega_1^\pm(k)
    =(S+s+\frac{(g_S-g_s)\mu_{\rm B}H}{2J}){\rm ch}2\theta_k
    -2\sqrt{Ss}{\rm cos}\frac{k}{2}{\rm sh}2\theta_k
    \pm(S-s)\mp\frac{(g_S+g_s)\mu_{\rm B}H}{2J}
   \nonumber\\
   &&\quad
    \equiv
    \omega_1(k)\pm(S-s)\mp\frac{(g_S+g_s)\mu_{\rm B}H}{2J},
   \nonumber\\
   &&
   \gamma_1(k)
    =2\sqrt{Ss}{\rm cos}\frac{k}{2}{\rm ch}2\theta_k
    -\Bigl[
      S+s+\frac{(g_S-g_s)\mu_{\rm B}H}{2J}
     \Bigr]
     {\rm sh}2\theta_k,
   \\
   &&
   \omega_0^\pm(k)
    =\Bigl[
      \Bigl(\sqrt{\frac{S}{s}}+\sqrt{\frac{s}{S}}\Bigr){\mit\Gamma}
     -2{\mit\Lambda}
     \Bigr]
     {\rm ch}2\theta_k
    +\Bigl[
      \Bigl(\sqrt{\frac{S}{s}}+\sqrt{\frac{s}{S}}\Bigr){\mit\Lambda}
     -2{\mit\Gamma}
     \Bigr]
     {\rm cos}\frac{k}{2}{\rm sh}2\theta_k
    \pm\Bigl(\sqrt{\frac{S}{s}}-\sqrt{\frac{s}{S}}\Bigr){\mit\Gamma},
   \nonumber\\
   &&
   \gamma_0(k)
    =\Bigl[
      2{\mit\Gamma}
     -\Bigl(\sqrt{\frac{S}{s}}+\sqrt{\frac{s}{S}}\Bigr){\mit\Lambda}
     \Bigr]
     {\rm cos}\frac{k}{2}{\rm ch}2\theta_k
    +\Bigl[
      2{\mit\Lambda}
     -\Bigl(\sqrt{\frac{S}{s}}+\sqrt{\frac{s}{S}}\Bigr){\mit\Gamma}
     \Bigr]
     {\rm sh}2\theta_k,
   \\
   &&
   V_0^{(1)}(k_1,k_2,k_3,k_4)=
   \Bigl({\rm cos}\frac{k_4-k_2}{2}+{\rm cos}\frac{k_3-k_1}{2}\Bigr)
   \bigl(
    {\rm ch}\theta_{k_1}{\rm sh}\theta_{k_2}
    {\rm ch}\theta_{k_3}{\rm sh}\theta_{k_4}
   +{\rm sh}\theta_{k_1}{\rm ch}\theta_{k_2}
    {\rm sh}\theta_{k_3}{\rm ch}\theta_{k_4}
   \bigr)
   \nonumber\\
   &&\quad
  +\Bigl({\rm cos}\frac{k_4-k_3}{2}+{\rm cos}\frac{k_2-k_1}{2}\Bigr)
   \bigl(
    {\rm ch}\theta_{k_1}{\rm ch}\theta_{k_2}
    {\rm sh}\theta_{k_3}{\rm sh}\theta_{k_4}
   +{\rm sh}\theta_{k_1}{\rm sh}\theta_{k_2}
    {\rm ch}\theta_{k_3}{\rm ch}\theta_{k_4}
   \bigr)
   \nonumber\\
   &&\quad
  -\sqrt{\frac{S}{s}}
   \Bigl[
   {\rm sh}\theta_{k_3}{\rm sh}\theta_{k_4}
   \Bigl(
    {\rm cos}\frac{k_1}{2}{\rm ch}\theta_{k_1}{\rm sh}\theta_{k_2}
   +{\rm cos}\frac{k_2}{2}{\rm sh}\theta_{k_1}{\rm ch}\theta_{k_2}
   \Bigr)
   +                 {\rm sh}\theta_{k_1}{\rm sh}\theta_{k_2}
   \Bigl(
    {\rm cos}\frac{k_3}{2}{\rm ch}\theta_{k_3}{\rm sh}\theta_{k_4}
   +{\rm cos}\frac{k_4}{2}{\rm sh}\theta_{k_3}{\rm ch}\theta_{k_4}
   \Bigr)
   \Bigr]
   \nonumber\\
   &&\quad
  -\sqrt{\frac{s}{S}}
   \Bigl[
    {\rm ch}\theta_{k_3}{\rm ch}\theta_{k_4}
   \Bigl(
    {\rm cos}\frac{k_1}{2}{\rm sh}\theta_{k_1}{\rm ch}\theta_{k_2}
   +{\rm cos}\frac{k_2}{2}{\rm ch}\theta_{k_1}{\rm sh}\theta_{k_2}
   \Bigr)
   +{\rm ch}\theta_{k_1}{\rm ch}\theta_{k_2}
   \Bigl(
    {\rm cos}\frac{k_3}{2}{\rm sh}\theta_{k_3}{\rm ch}\theta_{k_4}
   +{\rm cos}\frac{k_4}{2}{\rm ch}\theta_{k_3}{\rm sh}\theta_{k_4}
   \Bigr)
   \Bigr],
   \nonumber\\
   &&
   V_0^{(2)}(k_1,k_2,k_3,k_4)=
   \Bigl({\rm cos}\frac{k_4-k_2}{2}+{\rm cos}\frac{k_3-k_1}{2}\Bigr)
   \bigl(
    {\rm ch}\theta_{k_1}{\rm sh}\theta_{k_2}
    {\rm ch}\theta_{k_3}{\rm sh}\theta_{k_4}
   +{\rm sh}\theta_{k_1}{\rm ch}\theta_{k_2}
    {\rm sh}\theta_{k_3}{\rm ch}\theta_{k_4}
   \bigr)
   \nonumber\\
   &&\quad
  +\Bigl({\rm cos}\frac{k_4-k_3}{2}+{\rm cos}\frac{k_2-k_1}{2}\Bigr)
   \bigl(
    {\rm ch}\theta_{k_1}{\rm ch}\theta_{k_2}
    {\rm sh}\theta_{k_3}{\rm sh}\theta_{k_4}
   +{\rm sh}\theta_{k_1}{\rm sh}\theta_{k_2}
    {\rm ch}\theta_{k_3}{\rm ch}\theta_{k_4}
   \bigr)
   \nonumber\\
   &&\quad
  -\sqrt{\frac{S}{s}}
   \Bigl[
    {\rm ch}\theta_{k_3}{\rm ch}\theta_{k_4}
   \Bigl(
    {\rm cos}\frac{k_1}{2}{\rm sh}\theta_{k_1}{\rm ch}\theta_{k_2}
   +{\rm cos}\frac{k_2}{2}{\rm ch}\theta_{k_1}{\rm sh}\theta_{k_2}
   \Bigr)
   +{\rm ch}\theta_{k_1}{\rm ch}\theta_{k_2}
   \Bigl(
    {\rm cos}\frac{k_3}{2}{\rm sh}\theta_{k_3}{\rm ch}\theta_{k_4}
   +{\rm cos}\frac{k_4}{2}{\rm ch}\theta_{k_3}{\rm sh}\theta_{k_4}
   \Bigr)
   \Bigr]
   \nonumber\\
   &&\quad
  -\sqrt{\frac{s}{S}}
   \Bigl[
    {\rm sh}\theta_{k_3}{\rm sh}\theta_{k_4}
   \Bigl(
    {\rm cos}\frac{k_1}{2}{\rm ch}\theta_{k_1}{\rm sh}\theta_{k_2}
   +{\rm cos}\frac{k_2}{2}{\rm sh}\theta_{k_1}{\rm ch}\theta_{k_2}
   \Bigr)
   +{\rm sh}\theta_{k_1}{\rm sh}\theta_{k_2}
   \Bigl(
    {\rm cos}\frac{k_3}{2}{\rm ch}\theta_{k_3}{\rm sh}\theta_{k_4}
   +{\rm cos}\frac{k_4}{2}{\rm sh}\theta_{k_3}{\rm ch}\theta_{k_4}
   \Bigr)
   \Bigr],
   \nonumber\\
   &&
   V_0^{(3)}(k_1,k_2,k_3,k_4)=
   \Bigl({\rm cos}\frac{k_4-k_2}{2}+{\rm cos}\frac{k_3-k_1}{2}\Bigr)
   \bigl(
    {\rm ch}\theta_{k_1}{\rm ch}\theta_{k_2}
    {\rm ch}\theta_{k_3}{\rm ch}\theta_{k_4}
   +{\rm sh}\theta_{k_1}{\rm sh}\theta_{k_2}
    {\rm sh}\theta_{k_3}{\rm sh}\theta_{k_4}
   \bigr)
   \nonumber\\
   &&\quad
  +\Bigl({\rm cos}\frac{k_4-k_3}{2}+{\rm cos}\frac{k_2-k_1}{2}\Bigr)
   \bigl(
    {\rm ch}\theta_{k_1}{\rm sh}\theta_{k_2}
    {\rm sh}\theta_{k_3}{\rm ch}\theta_{k_4}
   +{\rm sh}\theta_{k_1}{\rm ch}\theta_{k_2}
    {\rm ch}\theta_{k_3}{\rm sh}\theta_{k_4}
   \bigr)
   \nonumber\\
   &&\quad
  -\sqrt{\frac{S}{s}}
   \Bigl[
    {\rm sh}\theta_{k_3}{\rm ch}\theta_{k_4}
   \Bigl(
    {\rm cos}\frac{k_1}{2}{\rm ch}\theta_{k_1}{\rm ch}\theta_{k_2}
   +{\rm cos}\frac{k_2}{2}{\rm sh}\theta_{k_1}{\rm sh}\theta_{k_2}
   \Bigr)
   +{\rm sh}\theta_{k_1}{\rm ch}\theta_{k_2}
   \Bigl(
    {\rm cos}\frac{k_3}{2}{\rm ch}\theta_{k_3}{\rm ch}\theta_{k_4}
   +{\rm cos}\frac{k_4}{2}{\rm sh}\theta_{k_3}{\rm sh}\theta_{k_4}
   \Bigr)
   \Bigr]
   \nonumber\\
   &&\quad
  -\sqrt{\frac{s}{S}}
   \Bigl[
    {\rm ch}\theta_{k_3}{\rm sh}\theta_{k_4}
   \Bigl(
    {\rm cos}\frac{k_1}{2}{\rm sh}\theta_{k_1}{\rm sh}\theta_{k_2}
   +{\rm cos}\frac{k_2}{2}{\rm ch}\theta_{k_1}{\rm ch}\theta_{k_2}
   \Bigr)
   +{\rm ch}\theta_{k_1}{\rm sh}\theta_{k_2}
   \Bigl(
    {\rm cos}\frac{k_3}{2}{\rm sh}\theta_{k_3}{\rm sh}\theta_{k_4}
   +{\rm cos}\frac{k_4}{2}{\rm ch}\theta_{k_3}{\rm ch}\theta_{k_4}
   \Bigr)
   \Bigr],
   \nonumber\\
   &&
   V_0^{(4)}(k_1,k_2,k_3,k_4)=
   \Bigl({\rm cos}\frac{k_4-k_2}{2}+{\rm cos}\frac{k_3-k_1}{2}\Bigr)
   \bigl(
    {\rm ch}\theta_{k_1}{\rm sh}\theta_{k_2}
    {\rm ch}\theta_{k_3}{\rm ch}\theta_{k_4}
   +{\rm sh}\theta_{k_1}{\rm ch}\theta_{k_2}
    {\rm sh}\theta_{k_3}{\rm sh}\theta_{k_4}
   \bigr)
   \nonumber\\
   &&\quad
  +\Bigl({\rm cos}\frac{k_4-k_3}{2}+{\rm cos}\frac{k_2-k_1}{2}\Bigr)
   \bigl(
    {\rm ch}\theta_{k_1}{\rm ch}\theta_{k_2}
    {\rm sh}\theta_{k_3}{\rm ch}\theta_{k_4}
   +{\rm sh}\theta_{k_1}{\rm sh}\theta_{k_2}
    {\rm ch}\theta_{k_3}{\rm sh}\theta_{k_4}
   \bigr)
   \nonumber\\
   &&\quad
  -\sqrt{\frac{S}{s}}
   \Bigl[
    {\rm ch}\theta_{k_3}{\rm sh}\theta_{k_4}
   \Bigl(
    {\rm cos}\frac{k_1}{2}{\rm sh}\theta_{k_1}{\rm ch}\theta_{k_2}
   +{\rm cos}\frac{k_2}{2}{\rm ch}\theta_{k_1}{\rm sh}\theta_{k_2}
   \Bigr)
   +{\rm ch}\theta_{k_1}{\rm ch}\theta_{k_2}
   \Bigl(
    {\rm cos}\frac{k_3}{2}{\rm sh}\theta_{k_3}{\rm sh}\theta_{k_4}
   +{\rm cos}\frac{k_4}{2}{\rm ch}\theta_{k_3}{\rm ch}\theta_{k_4}
   \Bigr)
   \Bigr]
   \nonumber\\
   &&\quad
  -\sqrt{\frac{s}{S}}
   \Bigl[
    {\rm sh}\theta_{k_3}{\rm ch}\theta_{k_4}
   \Bigl(
    {\rm cos}\frac{k_1}{2}{\rm ch}\theta_{k_1}{\rm sh}\theta_{k_2}
   +{\rm cos}\frac{k_2}{2}{\rm sh}\theta_{k_1}{\rm ch}\theta_{k_2}
   \Bigr)
   +{\rm sh}\theta_{k_1}{\rm sh}\theta_{k_2}
   \Bigl(
    {\rm cos}\frac{k_3}{2}{\rm ch}\theta_{k_3}{\rm ch}\theta_{k_4}
   +{\rm cos}\frac{k_4}{2}{\rm sh}\theta_{k_3}{\rm sh}\theta_{k_4}
   \Bigr)
   \Bigr],
   \nonumber\\
   &&
   V_0^{(5)}(k_1,k_2,k_3,k_4)=
   \Bigl({\rm cos}\frac{k_4-k_2}{2}+{\rm cos}\frac{k_3-k_1}{2}\Bigr)
   \bigl(
    {\rm ch}\theta_{k_1}{\rm sh}\theta_{k_2}
    {\rm ch}\theta_{k_3}{\rm ch}\theta_{k_4}
   +{\rm sh}\theta_{k_1}{\rm ch}\theta_{k_2}
    {\rm sh}\theta_{k_3}{\rm sh}\theta_{k_4}
   \bigr)
   \nonumber\\
   &&\quad
  +\Bigl({\rm cos}\frac{k_4-k_3}{2}+{\rm cos}\frac{k_2-k_1}{2}\Bigr)
   \bigl(
    {\rm ch}\theta_{k_1}{\rm ch}\theta_{k_2}
    {\rm sh}\theta_{k_3}{\rm ch}\theta_{k_4}
   +{\rm sh}\theta_{k_1}{\rm sh}\theta_{k_2}
    {\rm ch}\theta_{k_3}{\rm sh}\theta_{k_4}
   \bigr)
   \nonumber\\
   &&\quad
  -\sqrt{\frac{S}{s}}
   \Bigl[
    {\rm sh}\theta_{k_3}{\rm ch}\theta_{k_4}
   \Bigl(
    {\rm cos}\frac{k_1}{2}{\rm ch}\theta_{k_1}{\rm sh}\theta_{k_2}
   +{\rm cos}\frac{k_2}{2}{\rm sh}\theta_{k_1}{\rm ch}\theta_{k_2}
   \Bigr)
   +{\rm sh}\theta_{k_1}{\rm sh}\theta_{k_2}
   \Bigl(
    {\rm cos}\frac{k_3}{2}{\rm ch}\theta_{k_3}{\rm ch}\theta_{k_4}
   +{\rm cos}\frac{k_4}{2}{\rm sh}\theta_{k_3}{\rm sh}\theta_{k_4}
   \Bigr)
   \Bigr]
   \nonumber\\
   &&\quad
  -\sqrt{\frac{s}{S}}
   \Bigl[
    {\rm ch}\theta_{k_3}{\rm sh}\theta_{k_4}
   \Bigl(
    {\rm cos}\frac{k_1}{2}{\rm sh}\theta_{k_1}{\rm ch}\theta_{k_2}
   +{\rm cos}\frac{k_2}{2}{\rm ch}\theta_{k_1}{\rm sh}\theta_{k_2}
   \Bigr)
   +{\rm ch}\theta_{k_1}{\rm ch}\theta_{k_2}
   \Bigl(
    {\rm cos}\frac{k_3}{2}{\rm sh}\theta_{k_3}{\rm sh}\theta_{k_4}
   +{\rm cos}\frac{k_4}{2}{\rm ch}\theta_{k_3}{\rm ch}\theta_{k_4}
   \Bigr)
   \Bigr],
   \nonumber\\
   &&
   V_0^{(6)}(k_1,k_2,k_3,k_4)=
   2\Bigl({\rm cos}\frac{k_4-k_3}{2}+{\rm cos}\frac{k_2-k_1}{2}\Bigr)
   \bigl(
    {\rm ch}\theta_{k_1}{\rm sh}\theta_{k_2}
    {\rm sh}\theta_{k_3}{\rm ch}\theta_{k_4}
   +{\rm sh}\theta_{k_1}{\rm ch}\theta_{k_2}
    {\rm ch}\theta_{k_3}{\rm sh}\theta_{k_4}
   \bigr)
   \nonumber\\
   &&\quad
  -\sqrt{\frac{S}{s}}
   \Bigl[
    {\rm ch}\theta_{k_3}{\rm sh}\theta_{k_4}
   \Bigl(
    {\rm cos}\frac{k_1}{2}{\rm sh}\theta_{k_1}{\rm sh}\theta_{k_2}
   +{\rm cos}\frac{k_2}{2}{\rm ch}\theta_{k_1}{\rm ch}\theta_{k_2}
   \Bigr)
   +{\rm ch}\theta_{k_1}{\rm ch}\theta_{k_2}
   \Bigl(
    {\rm cos}\frac{k_3}{2}{\rm sh}\theta_{k_3}{\rm sh}\theta_{k_4}
   +{\rm cos}\frac{k_4}{2}{\rm ch}\theta_{k_3}{\rm ch}\theta_{k_4}
   \Bigr)
   \Bigr]
   \nonumber\\
   &&\quad
  -\sqrt{\frac{s}{S}}
   \Bigl[
    {\rm sh}\theta_{k_3}{\rm ch}\theta_{k_4}
   \Bigl(
    {\rm cos}\frac{k_1}{2}{\rm ch}\theta_{k_1}{\rm ch}\theta_{k_2}
   +{\rm cos}\frac{k_2}{2}{\rm sh}\theta_{k_1}{\rm sh}\theta_{k_2}
   \Bigr)
   +{\rm sh}\theta_{k_1}{\rm ch}\theta_{k_2}
   \Bigl(
    {\rm cos}\frac{k_3}{2}{\rm ch}\theta_{k_3}{\rm ch}\theta_{k_4}
   +{\rm cos}\frac{k_4}{2}{\rm sh}\theta_{k_3}{\rm sh}\theta_{k_4}
   \Bigr)
   \Bigr].
   \nonumber\\
   &&\ 
\end{eqnarray}
$\theta_k$ is determined so as to diagonalize the linear spin-wave
Hamiltonian ${\cal H}_1$, that is, to satisfy $\gamma_1(k)=0$.
$\alpha_k^\dagger$ ($\beta_k^\dagger$) creates spin waves of ferromagnetic
(antiferromagnetic) aspect.\cite{Y13610,Y211}
Minimizing the free energy under the effective condition of zero
staggered magnetization,\cite{Y157603} we obtain the optimum distribution
functions as
\begin{equation}
   \langle\alpha_k^\dagger\alpha_k\rangle\equiv\bar{n}_k^-,\ 
   \langle\beta_k^\dagger\beta_k\rangle\equiv\bar{n}_k^+;\ \ 
   \bar{n}_k^\sigma
  =\frac{1}
   {{\rm e}^{[J\omega_1^\sigma(k)-\mu{\rm ch}2\theta_{k}]/k_{\rm B}T}-1},
\end{equation}
where $\mu$ is determined through
\begin{equation}
   \sum_k\sum_\sigma
   \frac{S+s+(g_S-g_s)\mu_{\rm B}H/2J}{\omega_k}
   \bar{n}_k^\sigma
  =(S+s)Ne^{-J\omega_1^-(0)/k_{\rm B}T}.
\end{equation}

\section{Magnon Series Expansion of the Relaxation Rate}\label{A:T1}

   Considering the significant difference between the nuclear and
electronic energy scales and assuming the Fourier components of the
coupling constants to have little momentum dependence \cite{F433} as
$\sum_n e^{{\rm i}kn}A_n^\tau\equiv A_k^\tau\simeq A^\tau$ and
$\sum_n e^{{\rm i}kn}B_n^\tau\equiv B_k^\tau\simeq B^\tau$ ($\tau=-,z$),
we obtain the Raman and three-magnon relaxation rates as
\begin{eqnarray}
   &&
   \frac{1}{T_1^{(2)}}
   \simeq\frac{2(g\mu_{\rm B}\hbar\gamma_{\rm N}B^z)^2}{\hbar JN}
    \sum_{k_1}\sum_{\sigma=\pm}\sum_{\tau=\pm}
     W^{\sigma\sigma}(\tau k_2^{\sigma\sigma},k_1)^2
     \bar{n}_{k_1}^\sigma(\bar{n}_{k_2^{\sigma\sigma}}^\sigma+1)
     \left|
      \frac{{\rm d}\omega_1^\sigma(k)}{{\rm d}k}
     \right|_{k=k_2^{\sigma\sigma}}^{-1},
   \label{E:T1(2)}
   \\
   &&
   \frac{1}{T_1^{(3)}}
   \simeq\frac{(g\mu_{\rm B}\hbar\gamma_{\rm N}B^-)^2}{16\hbar SJN^2}
    \sum_{k_1,k_2}\sum_{\sigma=\pm}\sum_{\tau=\pm}
     2W^{\sigma\sigma\sigma}(\tau k_3^{\sigma\sigma\sigma},k_2,k_1)^2
     \bar{n}_{k_1}^\sigma\bar{n}_{k_2}^\sigma
     (\bar{n}_{k_3^{\sigma\sigma\sigma}}^{\sigma}+1)
     \left|
      \frac{{\rm d}\omega_1^\sigma(k)}{{\rm d}k}
     \right|_{k=k_3^{\sigma\sigma\sigma}}^{-1}
   \nonumber\\
   &&\qquad
    +W^{\sigma\bar{\sigma}\sigma}
     (k_1,k_2,\tau k_3^{\sigma\bar{\sigma}\sigma})^2
     \bar{n}_{k_1}^\sigma\bar{n}_{k_2}^{\bar{\sigma}}
     (\bar{n}_{k_3^{\sigma\bar{\sigma}\sigma}}^\sigma+1)
     \left|
      \frac{{\rm d}\omega_1^\sigma(k)}{{\rm d}k}
     \right|_{k=k_3^{\sigma\bar{\sigma}\sigma}}^{-1},
   \label{E:T1(3)}
\end{eqnarray}
with $\bar{\sigma}=-\sigma$,
where $k_2^{\sigma\sigma}$ and $k_3^{\sigma\sigma'\sigma}$ are determined
through
$\omega_{k_1}^\sigma-\omega_{k_2^{\sigma\sigma}}+\hbar\omega_{\rm N}/J=0$
and
$\omega_{k_1}^\sigma+\omega_{k_2}^{\sigma'}
-\omega_{k_3^{\sigma\sigma'\sigma}}^\sigma
+\sigma'\hbar\omega_{\rm N}/J=0$, respectively, and
\begin{eqnarray}
   &&
   W^{--}(k_1,k_2)
   =\frac{A^z}{B^z}{\rm ch}\theta_{k_1}{\rm ch}\theta_{k_2}
   -               {\rm sh}\theta_{k_1}{\rm sh}\theta_{k_2},\ \ 
   W^{++}(k_1,k_2)
   =\frac{A^z}{B^z}{\rm sh}\theta_{k_1}{\rm sh}\theta_{k_2}
   -               {\rm ch}\theta_{k_1}{\rm ch}\theta_{k_2},
   \\
   &&
   W^{---}(k_1,k_2,k_3)
   =\frac{A^-}{B^-}
    {\rm ch}\theta_{k_1}{\rm ch}\theta_{k_2}{\rm ch}\theta_{k_3}
   -\sqrt{\frac{S}{s}}
    {\rm sh}\theta_{k_1}{\rm sh}\theta_{k_2}{\rm sh}\theta_{k_3}
   -\frac{2SV_0^{(1)}(k_1,k_2,k_3,k_3+k_2-k_1)}
         {\omega_1^-(k_3+k_2-k_1)-\hbar\omega_{\rm N}/J}
   \nonumber\\
   &&\quad\times
    \Bigl(
     \frac{A^-}{B^-}{\rm ch}\theta_{k_3+k_2-k_1}
    -\sqrt{\frac{s}{S}}{\rm sh}\theta_{k_3+k_2-k_1}
    \Bigr)
   -\frac{ SV_0^{(4)}(k_1,k_2,k_3,k_3+k_2-k_1)}
         {\omega_1^+(k_3+k_2-k_1)+\hbar\omega_{\rm N}/J}
    \Bigl(
     \frac{A^-}{B^-}{\rm sh}\theta_{k_3+k_2-k_1}
    -\sqrt{\frac{s}{S}}{\rm ch}\theta_{k_3+k_2-k_1}
    \Bigr),
   \nonumber\\
   &&
   W^{+++}(k_1,k_2,k_3)
   =\frac{A^-}{B^-}
    {\rm sh}\theta_{k_1}{\rm sh}\theta_{k_2}{\rm sh}\theta_{k_3}
   -\sqrt{\frac{S}{s}}
    {\rm ch}\theta_{k_1}{\rm ch}\theta_{k_2}{\rm ch}\theta_{k_3}
   -\frac{SV_0^{(5)}(k_1,k_2,k_3,k_3+k_2-k_1)}
         {\omega_1^-(k_3+k_2-k_1)-\hbar\omega_{\rm N}/J}
   \nonumber\\
   &&\quad\times
    \Bigl(
     \frac{A^-}{B^-}{\rm ch}\theta_{k_3+k_2-k_1}
    -\sqrt{\frac{s}{S}}{\rm sh}\theta_{k_3+k_2-k_1}
    \Bigr)
   -\frac{ 2SV_0^{(2)}(k_1,k_2,k_3,k_3+k_2-k_1)}
         {\omega_1^+(k_3+k_2-k_1)+\hbar\omega_{\rm N}/J}
    \Bigl(
     \frac{A^-}{B^-}{\rm sh}\theta_{k_3+k_2-k_1}
    -\sqrt{\frac{s}{S}}{\rm ch}\theta_{k_3+k_2-k_1}
    \Bigr),
   \nonumber\\
   &&
   W^{+-+}(k_1,k_2,k_3)
   =2\Bigl(
     \frac{A^-}{B^-}
     {\rm sh}\theta_{k_1}{\rm ch}\theta_{k_2}{\rm sh}\theta_{k_3}
    -\sqrt{\frac{S}{s}}
     {\rm ch}\theta_{k_1}{\rm sh}\theta_{k_2}{\rm ch}\theta_{k_3}
     \Bigr)
   -\frac{ SV_0^{(3)}(k_3+k_2-k_1,k_3,k_2,k_1)}
         {\omega_1^-(k_3+k_2-k_1)-\hbar\omega_{\rm N}/J}
   \nonumber\\
   &&\quad\times
    \Bigl(
     \frac{A^-}{B^-}{\rm ch}\theta_{k_3+k_2-k_1}
    -\sqrt{\frac{s}{S}}{\rm sh}\theta_{k_3+k_2-k_1}
    \Bigr)
   -\frac{2SV_0^{(5)}(k_3,k_3+k_2-k_1,k_1,k_2)}
         {\omega_1^+(k_3+k_2-k_1)+\hbar\omega_{\rm N}/J}
    \Bigl(
     \frac{A^-}{B^-}{\rm sh}\theta_{k_3+k_2-k_1}
    -\sqrt{\frac{s}{S}}{\rm ch}\theta_{k_3+k_2-k_1}
    \Bigr).
   \nonumber\\
   &&
   W^{-+-}(k_1,k_2,k_3)
   =2\Bigl(
     \frac{A^-}{B^-}
     {\rm ch}\theta_{k_1}{\rm sh}\theta_{k_2}{\rm ch}\theta_{k_3}
    -\sqrt{\frac{S}{s}}
     {\rm sh}\theta_{k_1}{\rm ch}\theta_{k_2}{\rm sh}\theta_{k_3}
     \Bigr)
   -\frac{ 2SV_0^{(4)}(k_3,k_3+k_2-k_1,k_1,k_2)}
         {\omega_1^-(k_3+k_2-k_1)-\hbar\omega_{\rm N}/J}
   \nonumber\\
   &&\quad\times
    \Bigl(
     \frac{A^-}{B^-}{\rm ch}\theta_{k_3+k_2-k_1}
    -\sqrt{\frac{s}{S}}{\rm sh}\theta_{k_3+k_2-k_1}
    \Bigr)
   -\frac{SV_0^{(3)}(k_1,k_2,k_3,k_3+k_2-k_1)}
         {\omega_1^+(k_3+k_2-k_1)+\hbar\omega_{\rm N}/J}
    \Bigl(
     \frac{A^-}{B^-}{\rm sh}\theta_{k_3+k_2-k_1}
    -\sqrt{\frac{s}{S}}{\rm ch}\theta_{k_3+k_2-k_1}
    \Bigr).
   \nonumber\\
   &&\ 
\end{eqnarray}
\end{appendix}
\end{widetext}

\end{document}